# Optimal Supervisory Control Synthesis

Hassane ALLA
Hassane.Alla@inpg.fr
*Laboratoire GIPSA-Lab de Grenoble,*
*BP 46, 38402 St Martin d'Heres Cedex France*

**Abstract**: The place invariant method is well known as an elegant way to construct a Petri net controller. It is possible to use the constraint for preventing forbidden states. But in general case, the number forbidden states can be very large giving a great number of control places. In this paper is presented a systematic method to reduce the size and the number of constraints. This method is applicable for safe and conservative Petri nets giving a maximally permissive controller.

**Résumé**: La méthode des invariants est bien connus comme un moyen élégant pour synthétiser un contrôleur à partir d'un réseau de Petri. Il est possible d'utiliser les contraintes qui permettent d'empêcher d'aller ers les états interdits. Cependant dans le cas général, le nombre d'états interdits peut être très grand donnant ainsi un nombre très grand de places de contrôle. Cette méthode est applicable pour les réseaux de Petri conservatifs, et saufs donnant un contrôleur maximal permissif.

**Mots Clés**: contrôleur, invariant de marquage, réseau de Petri, états interdits, contrainte linéaire.

## 1. Introduction

Petri Nets (PN) is an appropriate and useful tool for the study of Discreet-Event Systems (DES) because of their modeling power and their mathematical properties. One of the advantages of PN models in DES control synthesis is the use of a PN marking as a distributed representation of the system state. For control goals represented as state avoidance problems, this often allows a decomposition of the control synthesis problem into a series of smaller problems which are more efficiently solved. In the last decade, the research in the field of controller synthesis of DES became one of the most active domains [1],[2],[3],[4].

In this paper we present an efficient method for synthesizing a controller to solve forbidden state problems for a particular class of DES's. We consider control specification which can be stated in the PN model. Synchronous composition between model of system and model of specification by existence the uncontrollable transitions, produce the forbidden states.

The forbidden state problem for a DES was introduced in a paper by Ramadge and Wonhom in which they demonstrated several fundamental properties of the solution [?]. The main limitation of such approach is the lack of structure in controlled automata and the large number of states of the related state transition structures.

Li and Wonham [??] have presented an algorithm which calculates the optimal solution for the systems modeled by Petri nets whose uncontrollable sub nets are loop-free. The controller has to solve on-line at each step linear integer programs. The approach will be difficult to apply to the grand systems because of computational complexity. A method for the computation of a maximally permissive controller using theory of region is presented in Ghaffari and Xie []. The main idea is to reduce the reachability graph to the set of admissible markings, i.e., markings that respect the specifications. Control places are added to the original network, so that the closed loop plant respecting the specifications. The theory of region is used on the reachability graph for the computation of the control places. For the computation of control places, a system of linear inequalities must be resolved. The solution of the system must be an integer positive solution. The advantage of this method is that gives a maximally permissive solution of the control problem, but the computation method is difficult and does not guarantee a solution. In this method, for each forbidden state, we must compute a control place. Then there is many control places and it increases the complexity of system. There is many of research for solving the problem of forbidden states [],[],[],[]. Each of them proposed the method for solving this problem. Most relevant to our work are [5],[6]. In [5], author presented a method for construct a optimal controller for the system modeled by Grafcet. The method is based on the equivalence between the set of forbidden states and the set of linear constraints deducted from it. The use of the Yamalidou technique in [5],[6] allows constructing a set of control places which represent the optimal controller. However the number of control places is equal to the number of constraints and can become very important. Intuitively, we realized that the PN solution is not unique from the point of view of the structure, but the reachable space (graph of the markings) is naturally unique. The problem comes from all the linear constraints which can be simplified by holding structural properties of the PN. In this paper is proposed a systematic method of reduction of the number of constraints. The minimal solution is not unique in the general case, but a method to choose the best solution is presented. This approach is based on the hypothesis that the PN models are safe and conservative. This paper is organized as below:

In the second section will be presented the fundamental definitions then how we can construct the linear constraints from forbidden states. Later, in the fourth section, the method to reduce the number and the size of the constraints will be presented. At the end, this method will be explained through an example.

## 2. Fundamental definitions

In this paper it is supposed that reader knows the bases of the Petri net paradigm and we present only the notions used in this paper. For more details, see the book in (David and Alla, 2005). A PN is presented by a 4-uplet $N = \{P, T, Pre, Post\}$ where $P$ is the set of places, $T$ the set of transitions, $Pre$: $P \times T \rightarrow N$ is the pre-incidence function that defines weighted arcs from places to transitions. $Post$: $T \times P \rightarrow N$ is the post-incidence function that defines weighted arcs from transitions to places. $C = Post - Pre$ is the incidence matrix.

The reachability graph is an automata $\mathbb{R} = \{M, \Sigma, \delta, m_0\}$ where $M$ is the states set, $\Sigma$ is the events set, $\delta: M \times \Sigma \rightarrow M$ the evolution function, $m_0$ is the initial state. This graph corresponds to all the possible evolutions of the PN. The reachability graph consists of nodes which correspond to the accessible markings, and of arcs to the firing of transitions. In this paper, we use the word state instead of marking, represented by all the marked places.

**Note 1**: The corresponding state to the marking $M_i$ will be noted as following:

$$M_i = (P_{i1} \ldots P_{ij} \ldots P_{in}) \mid \forall j \in \{1, \ldots, k\} \ m(P_{ij}) = 1 \ \& \ m(P_l) = 0 \ \text{for other places}$$

❑

For example the marking $M_1 = [0, 1, 1, 0, 0, 1, 0, 0, 0]$ will be presented as state $M_1 = P_2 P_3 P_6$.

In the reachability graph, there are some of states: the authorized state, the forbidden state and the not reachable state. Among the forbidden states a particular and important subset is constituted by the border forbidden states, it is denoted by the set $\mathcal{M}_B$. These states are such that all the input transitions are controllable. In this paper we use the definition that presented in (Toma 2005).

**Definition 1:** The set of forbidden states is noted $\mathcal{M}_F$ and is expressed:

$$\mathcal{M}_F = \{m \in M \mid \exists w \in L_R \setminus L_{Rd} \text{ and } m = \delta(m_0, w)\}$$

i.e., a state is forbidden if it is a reachable state and if it violates the specifications.

❑

**Definition 2:** The set of dangerous states is:

$$\mathcal{M}_D = \{m \in M \mid \exists w \in \Sigma_u^* \text{ and } \delta(m, w) \in \mathcal{M}_F\}$$

i.e., a dangerous state is a reachable marking from which there is at least a sequence of uncontrollable transitions who leads to a forbidden state. We will consider, also, that a forbidden state is a dangerous state.

❑

**Definition 3:** The set $\mathcal{M}_A$ of admissible states is the greatest set of reachable states so as:

- $\mathcal{M}_A \cap \mathcal{M}_D = \varnothing$
- If $m \in \mathcal{M}_A$ and $\delta(m, e) \in \mathcal{M}_F$ than $e \in \Sigma_c$, i.e., the passage from an admissible state to a dangerous state is made by the firing of a controllable transition.

Furthermore, $\mathcal{M}_A$ is the most permissive behavior respecting the specifications.

❑

**Definition 4**: The set of border states is:

$$\mathcal{M}_B = \{m \in \mathcal{M}_D \mid \exists\, e \in \Sigma_c \wedge \exists\, m_a \in \mathcal{M}_A,\ \text{s.t.}\ \delta(m_a, e) = m\}$$

i.e., the set of dangerous states which are reached by the firing of a controllable transition from an admissible state.

❑

**Definition 5:** A controller is **maximally permissive** (MPC) if all the admissible markings of $\mathcal{M}_A$ are reachable under supervision and all the firings of a transition, which cause the evolution of the plant from an admissible state to a non-admissible one, are inhibited.

❑

In this paper, an important definition will be also used; it concerns the notion of over-state.

**Definition 6**: Let be $M_i = P_{i1}\, P_{i2}\, \ldots P_{in}$ an admissible state. $M_j = P_{j1}\, P_{j2}\, \ldots P_{jm}$ is an over - state of $M_i$ if:

1) $\forall\, P_{jk} \in M_j\ \exists\ P_{ik} \in M_i\ \text{such as}\ P_{jk} = P_{ik} = 1$

2) $\forall\, P_{jl},\ l \notin [1, \ldots, m]\ M_j(P_{jl}) = 0\ \text{or}\ 1$

❑

For example state $M_1 = P_1 P_4 P_7$ is an over – state of $M_2 = P_1 P_4 P_7 P_{10}$.

## 3. From Forbidden States to Linear Constraints

The modelling approach presented in this paper, is the classic approach developed by Ramadge and Wonham (1987a, b, 1988). The model of the process is synchronized with the specifications giving the closed loop desired functioning. The existence of uncontrollable transition often leads to the existence of forbidden states including the border forbidden states. They are systematically determined by using the Kumar algorithm (Kumar and Holloway, 1996). We consider here that the PNs models of both process and specifications are safe, but this does not mean that the marking of the control places is Boolean. Let $f_i$ be a border forbidden state from $\mathcal{F}$ and $\{P_{i1} P_{i2} P_{i3} \ldots P_{in}\}$ be all the marked places which correspond to it. From the forbidden states[1], linear constraints can be constructed (Kattan, 2004; Toma et al, 2003). The linear constraint for the forbidden state $f_i$ is given by Equation (1).

$$\sum_{k=1}^{n} m_{ik} \leq n - 1 \qquad (1)$$

Where $n$ is the number of marked place of $f_i$, and $m_{ik}$ is the Boolean marking of place $P_{ik}$ in state $f_i$. For example: If the state $(P_2P_5P_7)$ is a forbidden state, it can be avoided by using Equation (2)

$$m_2 + m_5 + m_7 \leq 2 \qquad (2)$$

In this paper the forbidden state and its constraint is presented as below:

$$f_i = \{P_{i1}P_{i2}P_{i3} \ldots P_{in}\}$$
$$c_i = \{P_{i1}P_{i2}P_{i3} \ldots P_{in}, n\text{-}1\} \qquad (3)$$

The last term in $c_i$ corresponds to the number of marked places in the forbidden state minus 1, it is called the bound. This is true only at the beginning.

## 4. Reduction of the Size and Number of Constraints

In real systems, the number of forbidden states can be very large. Knowing that for each forbidden state, one control place is added, the complexity of the controller can become unmanageable. For this reason, in this paper we propose a systematic method to reduce the size and the number of the constraints.

The PNs models are assumed to be safe and conservative; then it is impossible for two places belonging to the same place invariant, to be marked simultaneously. This basic idea will be used for the simplification of the constraints.

**Property 1:** Let be $\{(P_1P_{i1} \ldots P_{i(n-1)}),\ldots,(P_r P_{i1} \ldots P_{i(n-1)})\}$ $r$ forbidden states in $M_F$.

$$m_1 + m_2 + \ldots + m_r = 1 \qquad (4)$$

The $r$ constraints are equivalent to one constraint as below:
$$m_1 + m_{i1} + \ldots + m_{i(n-1)} \leq n\text{-}1$$
$$m_2 + m_{i1} + \ldots + m_{i(n-1)} \leq n\text{-}1$$
$$\ldots$$
$$m_r + m_{i1} + \ldots + m_{i(n-1)} \leq n\text{-}1$$
$$\Longleftrightarrow$$
$$m_{i1} + \ldots + m_{i(n-1)} \leq n\text{-}2$$

Where $n$ is the number of marked place.

---

[1] Where there is no ambiguity, the word *border* will be omitted.

**Proof:** *necessary Condition*:
The sum of all constraints gives the constraint follow:
$(m_1+m_{i1}+\ldots+m_{i(n-1)}) + (m_2+m_{i1}+\ldots+m_{i(n-1)}) + \ldots + (m_r+m_{i1}+\ldots+m_{i(n-1)}) \leq r(n-1)$
(5)

(4), (5) $\Rightarrow$
$1+r(m_{i1}+\ldots+m_{i(n-1)}) \leq r(n-1) \quad \Rightarrow$
$m_{i1}+\ldots+m_{i(n-1)} \leq n-1-1/r \quad (m_i:\text{Integer number}) \Rightarrow$
$m_{i1}+\ldots+m_{i(n-1)} \leq n-2$

*Sufficient Condition*:
$m_{i1}+\ldots+m_{i(n-1)} \leq n-2$
$\forall\, k \in \{1,\ldots,r\}\ m_k \leq 1 \Rightarrow m_k + m_{i1}+\ldots+m_{i(n-1)} \leq n-1$

❏

Let be $\mathcal{F} = \{(P_1P_4P_7), (P_1P_4P_8), (P_1P_4P_9)\}$ set of a forbidden states and $m_7+m_8+m_9= 1$, thus the over-state $(P_1P_4)$ is forbidden state, without attention to marked place of the set $\{P_7, P_8, P_9\}$, or it is possible to use instead of three constraints only one constraint as below:

$m_1+m_4+m_7 \leq 2$
$m_1+m_4+m_8 \leq 2 \quad \xleftarrow{(m_7+m_8+m_9=1)}\rightarrow \quad m_1+m_4 \leq 1$
$m_1+m_4+m_9 \leq 2$

*NOTE*: it is possible to use property 1 more than one time.

❏

For example, if the set $\mathcal{F}$ is as below:

$\mathcal{F} = \{(P_1P_4P_7P_{10})\ ,\ (P_1P_4P_7P_{11})\ ,\ (P_1P_4P_8P_{10})\ ,\ (P_1P_4P_8P_{11}),\ (P_1P_4P_9P_{10}),\ (P_1P_4P_9P_{11})\}$
and
$m_{10} + m_{11}= 1$
$m_7+m_8+m_9= 1$

After the first use of Property 1 for the invariant between $P_{10}$ and $P_{11}$ the result is: $\{(P_1P_4P_7), (P_1P_4P_8), (P_1P_4P_9)\}$. Now it is possible to use Property 1 for the other invariant. The final result will be the over–state $(P_1P_4)$.

Let us consider another set of forbidden states as below;

$\mathcal{F}' = \{(P_1P_4P_7P_{10}), (P_1P_4P_7P_{11}), (P_1P_4P_8P_{10}), (P_1P_4P_8P_{11})\}$

At the end of first simplification the result is: $\{(P_1P_4P_7), (P_1P_4P_8)\}$ and it is not possible to use Property 1, but is it possible to reduce the number of constraint? The answer is yes. Because of the invariant between $P_7$, $P_8$ and $P_9$, the states $P_7$ and $P_8$ cannot be marked in the same time. Thus both constraints can decrease in a constraint as below:

$$m_1 + m_4 + m_7 + m_8 \leq 2 \qquad (6)$$

The property discussed above, is formalized in property 2:

*Property 2:* Let be $C = \{(P_1 P_{i1} \ldots P_{i(n-1)}, k), (P_2 P_{i1} \ldots P_{i(n-1)}, k), \ldots, (P_r P_{i1} \ldots P_{i(n-1)}, k)\}$ the equivalent constraints to forbidden states of marking graph and

$$m_1 + m_2 + \ldots + m_r \leq 1 \qquad (7)$$

The $r$ constraints are equivalent to one constraint as below:

$$m_1 + m_{i1} + \ldots + m_{i(n-1)} \leq k$$
$$m_2 + m_{i1} + \ldots + m_{i(n-1)} \leq k$$
$$\ldots$$
$$m_r + m_{i1} + \ldots + m_{i(n-1)} \leq k$$
$$\Longleftrightarrow$$
$$m_1 + m_2 + \ldots + m_r + m_{i1} + \ldots + m_{i(n-1)} \leq k$$

Where $n$ is the number of marked place.

*Proof: necessary Condition*:
$$\forall j \in \{1,\ldots,r\} \quad m_j + m_{i1} + \ldots + m_{i(n-1)} \leq k \qquad (8) \quad \text{and} \quad m_1 + m_2 + \ldots + m_r \leq 1$$
$$\Longrightarrow$$
$$(m_1 + m_2 + \ldots + m_r) + m_{i1} + \ldots + m_{i(n-1)} \leq k+1$$

We are going to show that limit $k+1$ is never reached. If not, it is necessary that $m_1 + m_2 + \ldots + m_r = 1$ and $m_{i1} + \ldots + m_{i(n-1)} = k$. But if $m_{i1} + \ldots + m_{i(n-1)} = k$, thus because of the constraints (8) it is necessary that for $\forall j \in \{1,\ldots,r\} \quad m_j = 0$. Then $m_1 + m_2 + \ldots + m_r = 0$. Thus limit $k+1$ never is reached.

*Sufficient Condition*:
$(m_1 + m_2 + \ldots + m_r) + m_{i1} + \ldots + m_{i(n-1)} \leq k$
and $\forall i \in \{1,\ldots,r\} \quad m_i \geq 0$
Then
 $\forall i \in \{1,\ldots,r\} \quad m_i + m_{i1} + \ldots + m_{i(n-1)} \leq k$
❑

*NOTE*: While exploiting this property, constraints are used instead of states since the bound may change according to the simplifications.
❑

**4.1 Unreachable states**

The objective of this approach is reduction the size and the numbers of the constraints. The final result is simpler if Property 1 is used. For this reason, our idea is to add the unreachable states in order to increase the space of forbidden states.

**Definition 7**: The unreachable states are the states that are not accessible from the initial state of the PN model or are reachable from border forbidden states.
❑

If these states help to reduce constraints, it is possible to use these states as forbidden states. But at the end, it is not necessary to choose a constraint which does not covered any forbidden state. This will be explained later in Section 5.

**4.2 Minimum of constraints**

In order to have the minimum of constraints, it is necessary to choose the simplest constraints containing all the forbidden states. To reach this goal it is necessary to write all the final constraints with the forbidden states. A method similar with the one proposed by Mac Cluskey for the reduction of the logical expressions is used for determining the best choice(Morris Mano 2001). Firstly, we must choose the simplified constraints which contain the forbidden states that are not in other constraints. Secondly for the forbidden states which are in two or several simplified constraints, it is needed to choose the constraint containing the greatest number of forbidden states which were not chosen, or the constraint which is simpler. In generally, there is not only one answer. This will be more explained in the following section.

## 5. EXAMPLE

To illustrate the ideas developed in this paper, let us consider the example which was presented in (Kattan 2004). A manufacturing system is composed of two independent machines $M1$ and $M2$, two transfer robots of the parts and one test bench where the final products are tested (Fig. 1).

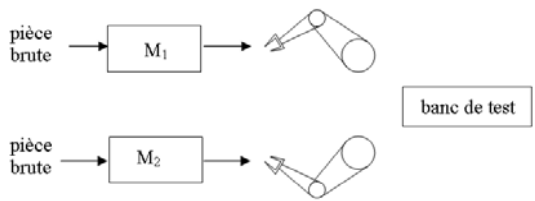

Fig. 1. Production line

Each machine has the following operating cycle: By occurrence of the event $c_i$, the machine starts working. When the work is finished (occurrence of event $f_i$) the produced part is transferred by the robot on the test bench, and a new cycle can be started again by occurrence of event $t_i$. There are two types of events in this system: the controllable events and uncontrollable events, only events $c_1$ and $c_2$ are controllable:

$$\Sigma_c = \{c_1, c_2\} \text{ et, } \Sigma_u = \{f_1, t_1, f_2, t_2\}.$$

The model PN for this system is given in Fig. 2.

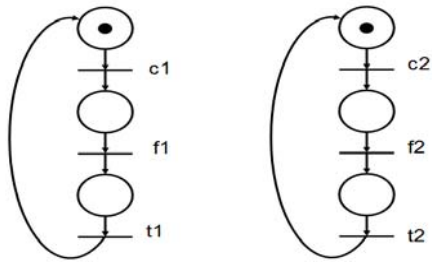 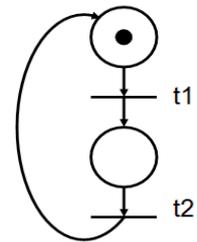

Fig. 2. PN process model.                    Fig. 3. PN specification model

The model of the specification of this system is presented in the Fig. 3. It is asked that firstly, the robot transfers the product of machine $M_1$ and then the product of machine $M_2$.

**5.1 Desired functioning in closed loop**

The determination of the forbidden states is made by the study of the controllability of the language of specification with regard to the language generated by the model of desired functioning in closed loop. In this example only transitions $c_1$ and $c_2$ are controllable and the other transitions are not controllable. The closed loop model of this system is illustrated in Fig. 4 and will be denoted as $R_d$. In this paper this model of PN is called Quasi-PN defined below.

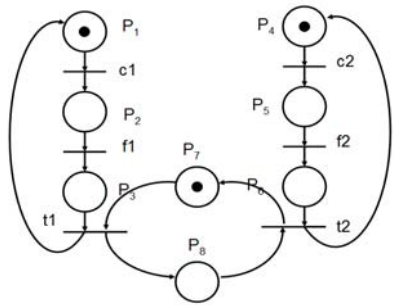

Fig. 4. Petri net model of closed loop system

**Definition 8:** A Quasi-PN is a PN which respects the following rules of firing:
    1) A controllable transition is firable in the same way as in a classical PN.
    2) An uncontrollable transition is firable if all its input places belonging to the process are marked
❏

The difference of reachable markings between these two models (PN and Quasi—PN) indicates the forbidden states.

**Definition 9**: If in from a marking, an uncontrollable transition is firable for the Quasi—PN model, but not for the real PN model, then this state is forbidden.
❏

To determine the forbidden states by this method, it is necessary to construct the reachability graph for both models.

### 5.2 Reachability graph of the functioning closed loop

The reachability graph is an identification tool of the behaviour of the PN. A node in the reachability Graph represents the reachable markings, and arcs characterize the possible passages between these markings. A marking corresponds to a combination of marked places. There are $2^N$ possible states for a safe PN with N places. But for a conservative PN, the number of states is smaller. Thanks to the method presented in (David and Alla, 2005), the set of possible states can be calculated. The number of state is determined as below:

$$N=\prod_{i=1}^{m}(n_i) \qquad (9)$$

Where $n_i$ is the number of places in each marking invariant and $m$ is the number of invariant.

From the reachability graph, the theory developed by Ramadge and Wonham allows to obtain the maximal permissive supervisor. If the language generated by the reachability graph is controllable, then the problem is resolved and the PN model $R_d$ constitutes the controller. But if it is not controllable, it is necessary to determine all the forbidden states.
The reachability graph of the real – PN and quasi – PN are presented in Fig. 5 and 6. The comparison of the two graphs, gives the forbidden states.

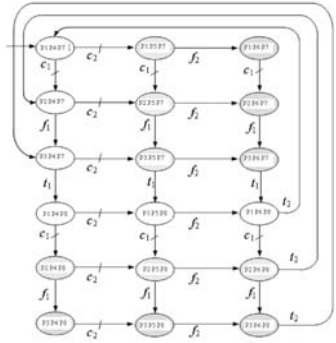 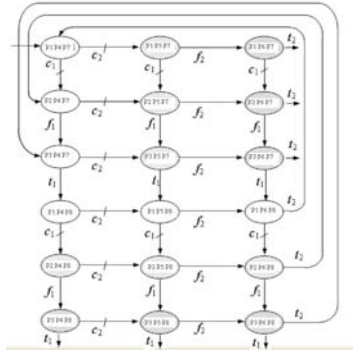

Fig. 5. Reachability graph of real – PN    Fig. 6. Reachability graph of quasi – PN

By using Definition 3 it is possible to determine the set of forbidden states as below:

$$E = \{(P_7P_6P_1), (P_7P_6P_2), (P_7P_6P_3), (P_8P_4P_3), (P_8P_5P_3), (P_8P_6P_3)\} \quad (10)$$

From these states, weakly forbidden states are deduced. By the technique presented in(Kattan 2004), it is possible to find all the forbidden and weakly forbidden states that are called dangerous states. To prevent from reaching the dangerous states, it is sufficient to forbid the border states. All the input transitions of these states are controllable.

For the example, the border states are:

$$\mathcal{F} = \{(P_7P_5P_1), (P_7P_5P_2), (P_7P_5P_3), (P_8P_4P_2), (P_8P_5P_2), (P_8P_6P_2)\} \quad (11)$$

From these 6 forbidden states, 6 constraints are deduced. We can now use the presented properties to reduce the size and the number of these constraints. For this, it is necessary to construct a set containing all the possible states and to indicate the type of each one.

In this example the marking invariants are obvious. For this example the number of possible states is calculated as below:

$$N = \prod_{i}^{m}(n_i) = 3*3*2 = 18 \quad (12)$$

| States | S | States | S | States | S |
|---|---|---|---|---|---|
| $P_7P_4P_1$ | 1 | $P_7P_4P_2$ | 1 | $P_7P_4P_3$ | 1 |
| $P_7P_5P_1$ | 0 | $P_7P_5P_2$ | 0 | $P_7P_5P_3$ | 0 |
| $P_7P_6P_1$ | Φ | $P_7P_6P_2$ | Φ | $P_7P_6P_3$ | Φ |
| $P_8P_4P_1$ | 1 | $P_8P_4P_2$ | 0 | $P_8P_4P_3$ | Φ |
| $P_8P_5P_1$ | 1 | $P_8P_5P_2$ | 0 | $P_8P_5P_3$ | Φ |
| $P_8P_6P_1$ | 1 | $P_8P_6P_2$ | 0 | $P_8P_6P_3$ | Φ |

1: admissible state  0: forbidden state  Φ: not reachable state

Fig. 7. All possible states

### 5.3 Use of Properties 1 and 2

In this example there are three invariants and Property 1 is used for all.

*I: $m_1+m_2+m_3=1$:*
Finding this invariant in Fig. 7 is easy, because in this case, all of the states are in one row. At the end it is obtained two over-states as below:

$(P_5P_7)$, it covers the states: $\{P_7P_5P_1, P_7P_5P_2, P_7P_5P_3\}$
$(P_6P_7)$, it covers the states: $\{P_7P_6P_1, P_7P_6P_2, P_7P_6P_3\}$
(13)

*II. $m_4+m_5+m_6=1$:*
By applying Property 1 for this invariant, the result is as below:

$(P_8P_2)$, covering $\{P_8P_4P_2, P_8P_5P_2, P_8P_6P_2\}$
$(P_8P_3)$, covering $\{P_8P_4P_3, P_8P_5P_3, P_8P_6P_3\}$   (14)

*III. $m_7+m_8=1$*
$(P_5P_2)$, covering $\{P_7P_5P_2, P_8P_5P_2\}$
$(P_6P_2)$, covering $\{P_7P_6P_2, P_8P_6P_2\}$
$(P_5P_3)$ , covering $\{P_7P_5P_3, P_8P_5P_3\}$
$(P_6P_3)$, covering $\{P_7P_6P_3, P_8P_6P_3\}$            (15)

The following table contains all the over states and indicates all the forbidden states which are covered by them. If may happen that a state is not covered, then it must be added in the final selection. It is not the case in this example (Fig. 8).

| Over- | The states which are covered | S |
|---|---|---|

| states | | |
|---|---|---|
| $P_5P_7$ | $P_7P_5P_1, P_7P_5P_2, P_7P_5P_3$ | 0 |
| $P_6P_7$ | $P_7P_6P_1, P_7P_6P_2, P_7P_6P_3$ | Φ |
| $P_8P_2$ | $P_8P_4P_2, P_8P_5P_2, P_8P_6P_2$ | 0 |
| $P_8P_3$ | $P_8P_4P_3, P_8P_5P_3, P_8P_6P_3$ | Φ |
| $P_5P_2$ | $P_7P_5P_2, P_8P_5P_2$ | 0 |
| $P_6P_2$ | $P_7P_6P_2, P_8P_6P_2$ | 0 |
| $P_5P_3$ | $P_7P_5P_3, P_8P_5P_3$ | 0 |
| $P_6P_3$ | $P_7P_6P_3, P_8P_6P_3$ | Φ |

Fig. 8. All states after first simplification

Here, Property 1 cannot be used longer, but in general case, it is possible to examine the table for new simplifications. Property 2 can now be used.

From now, we use the constraints instead of states or over-states since the bound may change according to the simplifications. There is two over-states or two constraint $\{(P_5P_7, 1), (P_6P_7, 1)\}$ which can be simplified by using Property 2. It is possible to use instead of two constraints only one constraint as bellow:

$$\{(P_5P_7, 1), (P_6P_7, 1)\} \Rightarrow (P_5 P_6P_7, 1) \quad (16)$$

The result after using Property 2 for all the constraints of Fig. 8 is illustrated in Fig. 9.

| Constraint | The states which are covered | S |
|---|---|---|
| $P_5 P_6 P_7, 1$ | $P_7P_5P_1, P_7P_5P_2, P_7P_5P_3, P_7P_6P_1, P_7P_6P_2, P_7P_6P_3$ | 0 |
| $P_8 P_2 P_3, 1$ | $P_8P_4P_2, P_8P_5P_2, P_8P_6P_2, P_8P_4P_3, P_8P_5P_3, P_8P_6P_3$ | 0 |
| $P_5 P_6 P_2 P_3, 1$ | $P_7P_5P_2, P_8P_5P_2, P_7P_6P_2, P_8P_6P_2, P_7P_5P_3, P_8P_5P_3, P_7P_6P_3, P_8P_6P_3$ | 0 |

Fig. 9. All states after the last simplification

### 5.4 Minimum of constraint for this example

At the end of this level, it is necessary to choose the minimum and simplest constraints, containing all the forbidden states. In general case, it is possible to have some choice. The difference between two choices is the number of arc of the control places. The best choice is determined after the calculation of the control places.

|  | $P_7P_5P_1$ | $P_7P_5P_2$ | $P_7P_5P_3$ | $P_8P_4P_2$ | $P_8P_5P_2$ | $P_8P_6P_2$ | choice |
|---|---|---|---|---|---|---|---|
| $P_5 P_6 P_7, 1$ | √ | √ | √ |  |  |  | √ |
| $P_8 P_2 P_3, 1$ |  |  |  | √ | √ | √ | √ |
| $P_5 P_6 P_2 P_3, 1$ |  | √ | √ |  | √ | √ |  |
|  | √ | √ | √ | √ | √ | √ |  |

Fig. 10. Minimum of constraints

At the end, the best solution contains two linear constraints(Fig. 10).

$$m_5+m_6+m_7 \leq 1$$

$$m_2+m_3+m_8 \leq 1 \qquad (17)$$

## 5.5 Calculation of control places

To calculate control places for every linear constraint, there is a systematic method given in (Yamalidou et al 1996). The incidence matrix of the PN $R_d$, corresponding to the model of functioning in closed loop is $W_p$. The synthesis approach consists in adding to the incidence matrix $W_p$ a matrix $W_c$ corresponding to the control places. The incidence matrix of controlled system $W$ will be constructed by adding a line to the matrix of incidence of controlled process $W_p$ for every control place.

$$W = \begin{bmatrix} W_p \\ W_c \end{bmatrix} \qquad (18)$$

It is then possible to calculate the matrix $W_c$ and its initial marking as bellow:

$$W_c = -L.W_P \qquad (19)$$

And

$$M_{cinit} = b - L.M_{pinit} \qquad (20)$$

The controller is known by the determination of his matrix of incidence $W_c$ and his initial marking $M_{cinit}$. By applying this result to the example, the result is:

$$Wc = \begin{bmatrix} -1 & 0 & 0 & 0 & 0 & 1 \\ 0 & 0 & 1 & -1 & 0 & 0 \end{bmatrix}$$

Fig. 11. The control matrix $W_c$

The initial marking of the control place is calculated:
  $m_{c1} = 1$;
  $m_{c2} = 0$;
The PN of the final controller is represented in Fig.12.

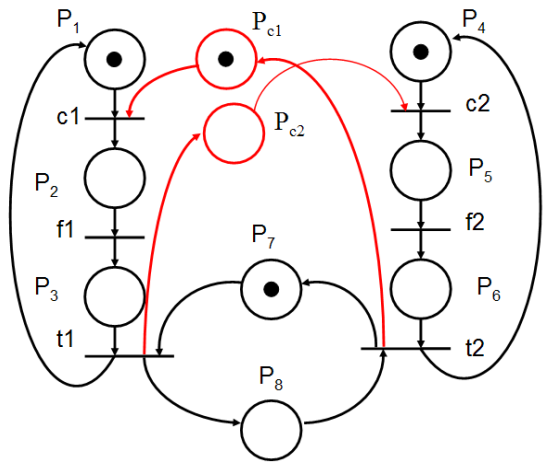

Fig. 12. Petri net model of closed loop system with control places

## 6. Problem of state explosion

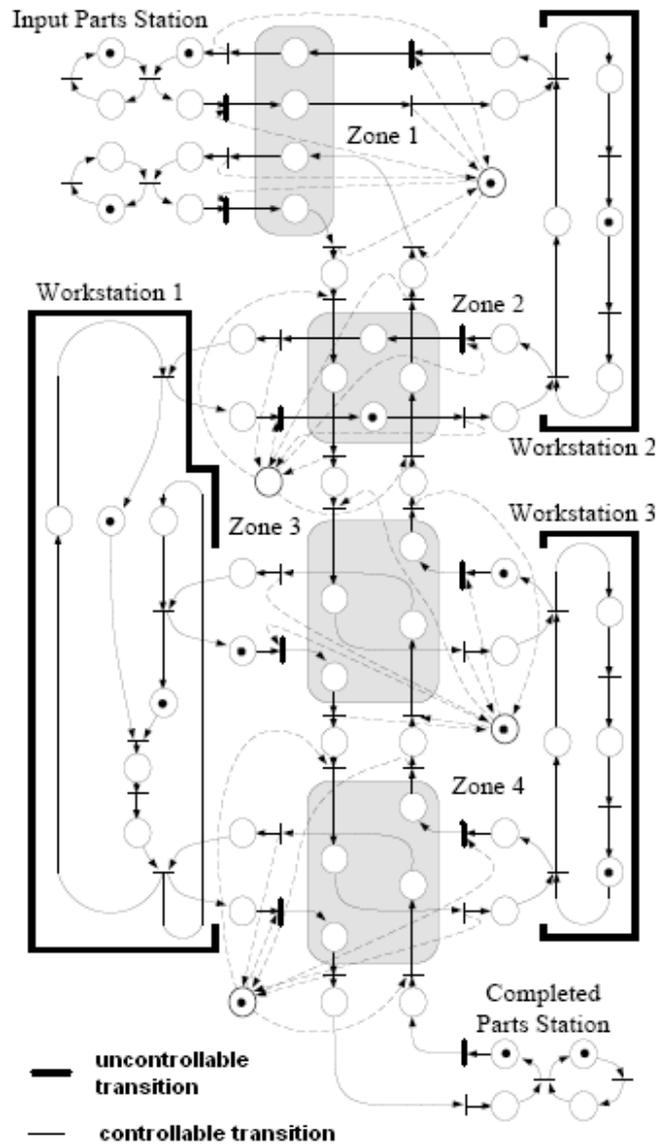

Fig. 13. The controlled Automated Guided Vehicle

In the large system, the number of possible states is enormous and by this method we need to large memory and long time for calculating the simpler controller. By example for a system AVG by two inputs and two outputs and 4 workstations and 4 critical zones, we have 30,965,760 states possible and computing the controller is very complicated. But it is obvious that the situation of output system don't effect on the situation of Wagon In zone 1. Therefore we can compute the controller for the zone 1 without attention to mark in the model of output system. The Petri Net model for controlled AVG is shown in Fig. 13.

In this example, supervisor is presented by the constraints. By Yamalidou's method, the supervisor is computed. But because of uncontrollable transition this supervisor doesn't work correctly. For resolve the problem of first zone we can neglect the model of workstation 1, 2, 3 and the Wagons that move out of zone 1. This model is presented in Fig. 14.

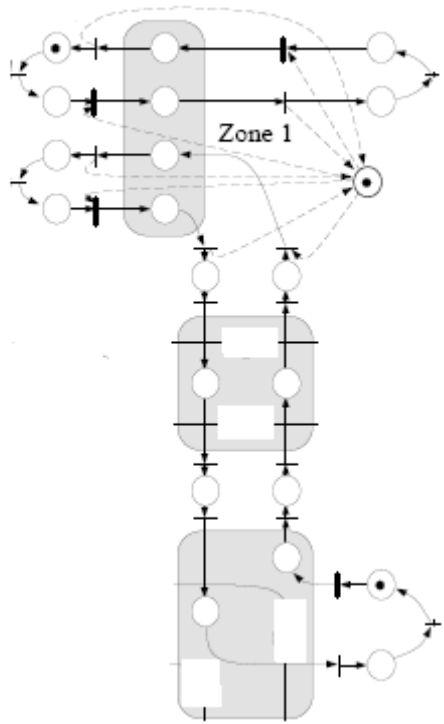

Fig. 14. The dependence invariants to zone 1

Now, we can compute the forbidden states and border forbidden states for this system and compute the controller by the method that presented in this paper.
**Remark:** The controller isn't always simpler controller, because some times we can construct one control place for two or more critical zone.

**Remark:** In supervisory idea the critical zone produced when we have an uncontrollable transition common between model of system and model of supervisor.

We must find all of this type of transitions and dependence invariants. Finding the invariants that depend to each critical zone is possible intuitively for the simple system, but in the large system we need to an algorithm to computing the dependence invariant for each critical zone. The result for this example is shown in Fig. 15.

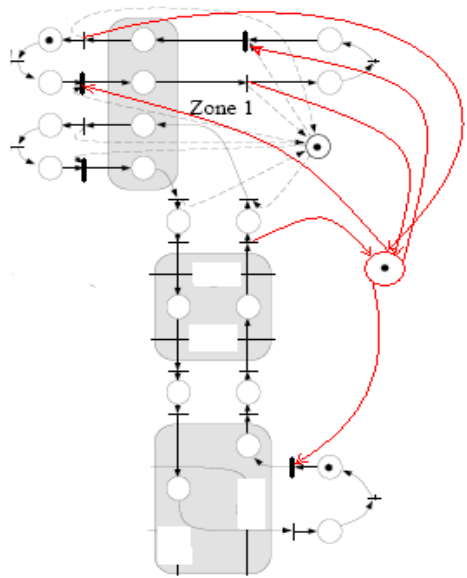

Fig. 15. The controller for zone 1

In this example we can apply the constraints for supervisory as the forbidden states in constructing the controller. This result is presented in Fig 16.

## 6. Conclusion
In this paper we have presented a systematic method to reduce the number of forbidden states or the number of equivalent linear constraints. The basic idea is to use the markings invariants of the PN to simplify these constraints. This is realized by using the not reachable states.

We obtain a structural solution easy to implement. The solution obtained by our approach gives the optimal controller. This optimality comes from the equivalence between all the admissible markings and all the linear constraints. The method that is presented in this paper is applicable for the systems modelled by Grafcet if the place of control is safe.

Our future work consists in: 1) establishing complete algorithms and, 2) developing this approach for the system that is modelled by safe PN model not necessarily conservative.

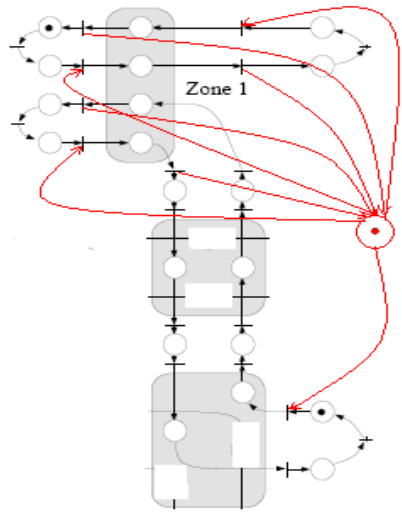

Fig. 16. The controlled system without supervisor for zone 1